# A Quantum Cryptographic Protocol with Detection of Compromised Server

D. Richard Kuhn

National Institute of Standards and Technology
Gaithersburg, MD  20899   USA

**Abstract** *This paper presents a hybrid cryptographic protocol, using quantum and classical resources, to generate a key for authentication and optionally for encryption in a network.  One or more trusted servers distribute streams of entangled photons to individual resources that seek to communicate. An important class of cheating by a compromised server will be detected.  Each resource shares a previously distributed secret key with the trusted server, and resources can communicate with the server using both classical and quantum channels.  Resources do not share secret keys with each other, so that the key distribution problem for the network is reduced from $O(n^2)$ to $O(n)$.  Some advantages of the protocol are that it avoids the requirement for timestamps used in classical protocols, guarantees that the trusted server cannot know the authentication key, can provide resistance to multiple photon attacks [Brassard et al., 1999; Felix et al., 2001] and can be used with BB84 [Bennett84] or other quantum key distribution protocols.*

# 1   Introduction

Controlling access to a large network of resources is one of the most common security problems.  Familiar examples of authentication include the process of supplying a password to gain access to a computer, or use of a personal identification number (PIN) with an automatic teller machine.   The user seeking authentication must provide some ticket that cannot be held by anyone else, either because user and system shared the secret at some point in the past, or both received the secret from some trusted third party with assurance that the communication was not intercepted.

Any pair of parties in a network should be able to communicate, but must be authorized to do so, which requires that their identities be authenticated.  The fundamental problem is how to authenticate resources to each other while minimizing the number of cryptographic keys that must be distributed and maintained, given the potential for $n(n-1)/2$ pairs of communicating resources.  Conventional solutions are typically based on authentication protocols such as Kerberos or public key schemes, which use trusted servers to grant authentication tickets or certificates to the communicating parties.   Less sophisticated examples include the use of simple passwords on a network.  Password or authentication key transmission may or may not be encrypted, depending on the level of risk.

This paper describes a solution based on a combination of quantum cryptography and a conventional secret key system (although a public key system could be used for the classical component as well).  A novel feature of this approach is that even the trusted server cannot know the contents of the authentication ticket.  Using quantum cryptography also avoids the need for timestamps and key expiration periods.

## 1.1   Protocol Description

This section describes the protocol (illustrated in Figure 1) under idealized conditions.  A later section discusses the impact of transmission losses, detection rates and other limiting factors of physical implementations.  We assume that each resource shares a secret key with a trusted server that an eavesdropper can read but not modify messages, and that resources can communicate with the trusted server over a classical and quantum channel.  For the discussion below, it is assumed that the trusted server can be, in fact, trusted.  A subsequent section describes a modification to the protocol that allows detection of a compromised server.

1. On the classical channel Alice sends a message to the trusted server, Tr, encrypted under Alice's secret key, indicating the party, Bob, that Alice seeks to communicate with. (A classical communication channel is suggested here, but the only requirement is that parties be able to communicate securely with the trusted server.  Any form of secure communication could be used.  Authentication between Alice and the trusted server is also required, and can be accomplished through a variety of existing classical protocols that are not described here. )
2. Using the secret keys shared with Alice and Bob, Tr sends to Alice and Bob the location, basis, and polarization of tamper detection bits.

3. On the quantum channels Tr sends a stream of *k* pairs of authentication key bits along with *d* pairs of randomly interspersed tamper detection bits. Each key bit is one half of an entangled pair of photons in the state $\Phi^+ = \frac{1}{\sqrt{2}}(|00\rangle + |11\rangle)$.

4. One photon of each pair goes to Alice and its twin to Bob. The tamper detection bit pairs are polarized randomly, according to a sequence of randomly selected bases. Each photon in a pair is polarized in the same direction as the other; one is sent to Alice and its twin to Bob.

5. Alice and Bob measure key photons according to a pre-determined basis, known to all communicating parties, and tamper detection photons according to the sequence of bases received from Tr, producing a sequence of authentication key bits and tamper detection bits. That is, a measurement result of $|0\rangle$ is treated as a key bit of 0 and a measurement result of $|1\rangle$ is treated as a key bit of 1.

   *Key bits measurement:* Since the key bits are entangled, Bob will observe the same measurement seen by Alice.

   *Tamper detection bits measurement:* With zero transmission loss and perfect detection, the tamper detection bits will match Tr's message with 100% accuracy. If an eavesdropper, Eve, has read the message the error rate for tamper detection bits will be 25%, since she has a 50% chance of guessing the correct basis, and a 50% chance that Alice and Bob will measure the correct polarization even if Eve chooses the wrong basis. In a practical implementation, the error threshold for tamper detection bits should be set as close to 0 as practical, for reasons discussed in a subsequent section. If the error rate for tamper detection bits exceeds the error threshold, the protocol is restarted.

6. To authenticate her identity to Bob, Alice sends to Bob the result of measuring the key bit sequence to provide confidence (with probability $1 - 2^{-k}$) that the message is from Alice. The authentication key effectively serves as a session password, which is sent in the clear. Note that Alice may send only a portion of the key bit sequence, sufficient to authenticate her identity, while retaining the rest to be used as a shared secret key. That is, the protocol can incorporate key distribution as well as authentication.

7. Bob compares his measurement of the photon stream received from Tr with the result sent by Alice. A perfect match authenticates Alice.

After step 6, Alice and Bob share a bit sequence resulting from their measurement of the key photons, and even Tr cannot know the bit sequence for the bits that were measured because the measurement result is not transmitted. Note also that after step 6, Eve will gain nothing by decrypting communications between the trusted server and Alice and Bob, because knowing the location of tamper detection bits is of no value after measurements are made on the key bits. This information needs to be protected for only a few seconds or milliseconds, making it possible – with sufficient key length – to resist attacks from even a quantum computer.

At the end of an exchange, some portion of Alice and Bob's shared bit sequence might be used as an encryption key as well, although doing so involves greater risk than using the bits as an authentication key because leaking partial information can make the key vulnerable. Privacy amplification techniques might be used to reduce Eve's information in this case [Bennett et al., 1995]. More on the potential for Eve guessing bit values is discussed in following sections.

## 1.2 Detecting a Compromised Server

The previous discussion assumed that the server could be trusted. However, if the server were compromised, it could send matching qubits to Alice and Bob, in place of entangled pairs. Since Alice and Bob measure their qubits individually, they would have no way of recognizing that the qubits received from Tr were not entangled, and therefore could not detect the fact that the server could have retained a copy of their bit streams. An addition to the protocol, using entanglement swapping [Zukowski et al., 1993; Cabello, 2000; Song, 2003], will prevent this attack, ensuring that the bit stream developed by Alice will not authenticate with Bob, to an arbitrarily high probability. The protocol proceeds as described previously, with Step 5 replaced as follows:

Step 5:
5a. For each qubit received from Tr, Alice prepares an entangled pair in a random Bell state, $|S\rangle$, where $|S\rangle$ is one of the conventional Bell states $\Phi^\pm = (|00\rangle \pm |11\rangle)/\sqrt{2}$ and $\Psi^\pm = (|01\rangle \pm |10\rangle)/\sqrt{2}$. If the server is operating as expected, the qubits received by Alice and Bob are in the state $\Phi^+_{AB} = \frac{1}{\sqrt{2}}(|0_k 0_l\rangle + |1_k 1_l\rangle)$, where qubit $Q_k$ is the qubit received by Alice and $Q_l$ is the qubit received by Bob.

5b. With the combined system $|S\rangle \otimes |\Phi^+_{AB}\rangle$ Alice executes a Bell basis measurement on $Q_k$ (which she received from Tr) and qubit $Q_j$ that she created. As a result of this measurement, Alice's qubit $Q_i$ becomes entangled with the qubit $Q_l$ that Bob received from Tr. Because she knows the initial state of her created qubits $Q_i$ and $Q_j$, her result from measuring $Q_j$ and $Q_k$, and the state of qubits sent from Tr, Alice can determine the state of the now entangled qubits $Q_i$ and $Q_l$. Using the state of $Q_i$ and $Q_j$, and the result of her measure of $Q_j$ and $Q_k$, the entangled state of $Q_i$ and $Q_l$ can be determined from Table 1 below:

|  | Result of measuring $Q_j Q_k$ | | | |
| --- | --- | --- | --- | --- |
| State of $Q_i Q_j$ | $\Phi^+$ | $\Phi^-$ | $\Psi^+$ | $\Psi^-$ |
| $\Phi^+$ | $\Phi^+$ | $\Phi^-$ | $\Psi^+$ | $\Psi^-$ |
| $\Phi^-$ | $\Phi^-$ | $\Phi^+$ | $\Psi^-$ | $\Psi^+$ |
| $\Psi^+$ | $\Psi^+$ | $\Psi^-$ | $\Phi^+$ | $\Phi^-$ |
| $\Psi^-$ | $\Psi^-$ | $\Psi^+$ | $\Phi^-$ | $\Phi^+$ |

**Table 1.** Entanglement of $Q_i$ and $Q_l$

For example, if $Q_i$ and $Q_j$ are created in state $\Phi^+$ and the result of her measurement on $Q_j$ and $Q_k$ is $\Phi^+$, then she knows that $Q_i$ and $Q_l$ will then be in state $\Phi^+$, or if $Q_j$ and $Q_k$ are in $\Psi^-$, then $Q_i$ and $Q_l$ must also be in $\Psi^-$.

5c. Alice now measures her qubit $Q_i$. When Bob measures his corresponding qubit, $Q_l$, his result will be either correlated or anti-correlated with Alice's result from measuring $Q_i$. Alice establishes her key bit accordingly:

| $Q_i Q_l$ state | $Q_i$ result | Key bit |
| --- | --- | --- |
| $\Phi^+$ or $\Phi^-$ | $|0\rangle$ | 0 |
| $\Phi^+$ or $\Phi^-$ | $|1\rangle$ | 1 |
| $\Psi^+$ or $\Psi^-$ | $|0\rangle$ | 1 |
| $\Psi^+$ or $\Psi^-$ | $|1\rangle$ | 0 |

5d. Bob waits to measure his qubits until a transmission is received from Alice. At this point, he measures and sets key bits as described in section 1.1. If the qubits received from Tr were in fact entangled, then Alice and Bob now share a common bit sequence not known to Tr or other parties.

Suppose now that Tr has been compromised and it attempts to record a copy of Alice and Bob's shared bit sequence by sending non-entangled random qubits to Alice and Bob, recording the sequence sent. In this case, the combined system at step 5b is $|S\rangle \otimes |x\rangle$, with $x$ either 0 or 1. Since Alice creates one of the four Bell states randomly, the Bell state measurement of $Q_j$ and $Q_k$ will result in a $\Phi$ or $\Psi$ state with equal likelihood, and the generated key bit will match or not

| State of $Q_i Q_j$ | $Q_k$ and $Q_l$ sent by Tr | Measurement $Q_j$ and $Q_k$ | Resulting key bit |
| --- | --- | --- | --- |
| $\Phi^\pm$ | $|x\rangle$ | $\Phi^\pm$ | $|x\rangle$ |
|  | $|x\rangle$ | $\Psi^\pm$ | $|x\rangle$ |
| $\Psi^\pm$ | $|x\rangle$ | $\Phi^\pm$ | $|\bar{x}\rangle$ |
|  | $|x\rangle$ | $\Psi^\pm$ | $|\bar{x}\rangle$ |

**Table 2.** Key bits with compromised server match Bob's key bit with probability .50. For example, if Alice generates $\Psi^+$ and Tr sends $|0\rangle$, then the combined state is $\Psi^+ \otimes |0\rangle$. If the result of measuring is $Q_j$ and $Q_k$ is $\Phi^\pm$ then $Q_l$ will be measured as $|1\rangle$ and Alice's key bit will be 1. Since the qubits sent by Tr are not entangled, the value of Bob's qubit is the 0 bit sent by the compromised server. So Alice's key bit value will match Bob's key bit value with only probability .50. Alice's authentication key will therefore be rejected with probability $1 - 2^{-n}$, for $n$ key bits. So the copy made by the compromised Tr could not be used for authentication.

Another server-based attack is possible using GHZ state qubits. If the server can create three qubits in GHZ state, retaining one and forwarding the others to Alice and Bob, it could develop a copy of their key bit stream. However, this attack is in a distinctly different risk category than the attack described previously. The attack using matched non-entangled qubits could conceivably be accomplished remotely, across a network, by modifying the software controlling the server. But the server could not be simply reprogrammed to generate GHZ state qubits, since more sophisticated equipment would be required. This attack would require physical access to the server, to install equipment capable of generating GHZ states, so it could be mitigated by conventional physical security.

Another possibility is that the compromised server could forge a message from Alice to Bob. Since the server has access to the original secret keys shared with each user, the modifications to step 5 do not prevent this attack, but they do prevent the server from copying a key used for encryption, rather than authentication.

## 2   Analysis of Security Properties

This section considers possible attacks against the protocol and examines parameters required for a desired level of security.

### 2.1   Intercept-resend attack

Suppose that Eve intercepts the photon stream going to either Alice or Bob, and resends. In this case, she must guess the basis for the tamper detection bits, guessing incorrectly 50% of the time. Alice (or Bob) will measure the tamper detection bits according to the basis sent by Tr. If the tamper detection bits have not been measured by Eve, then the polarization measured by Alice will agree with that sent by Tr 100% of the time and Alice will observe an error rate of 0. If the tamper detection bits have been measured by Eve, then Alice will observe an error rate of .25.

Guessing which bits are for tamper detection and which for the authentication key is not a feasible strategy. Tamper detection bits are interspersed randomly, so the chance of picking the correct $k$ key bits out of $k+d$ bits is $\binom{k+d}{k}^{-1}$, which will be extremely small for reasonable values of $k$ and $d$, where $k$ is the number of authentication key bits and $d$ the number of tamper detection bits. Eve could try guessing a subset of the bits, hoping to get all $k$ key bits without disturbing the tamper detection bits. The chance of this strategy succeeding for guessing a total of $g$ bits is a product of the probability of getting all $k$ key bits and the probability of disturbing a tamper detection bit:

$$\frac{\binom{k+d-k}{g-k}}{\binom{k+d}{g}} \cdot (0.75^{g-k})$$

$$= \frac{d!\,g!}{(k+d)!(g-k)!} \cdot (0.75^{g-k})$$

Eve has a tradeoff in that increasing the number of guessed bits, $g$, increases her chances by making it more likely to get all $k$ key bits, but decreases them by raising the probability that an error detection bit will be disturbed, thus revealing her presence. Overall, Eve's chances of success increase as more bits are guessed. Reading an extra bit will increase the left side of the product by a factor of

$$\frac{\frac{d!(g+1)!}{(k+d)!(g+1-k)!}}{\frac{d!\,g!}{(k+d)!(g-k)!}} = \frac{g+1}{g+1-k}$$

Since $g > k$, $\frac{g+1}{g+1-k} > 1$, but the right side will decrease by a factor of only 0.75 for each extra bit guessed. Therefore Eve's chances improve as long as

$\frac{g+1}{g+1-k} \cdot 0.75 > 1$, or up to a limit of $g < 4k - 1$. As shown below, this limit is not reached if the probability of falsifying an authentication token and the probability of evading detection of eavesdropping are balanced. The best strategy for Eve, then, is to measure all bits and hope that the measurement does not induce an error detectable by Alice and Bob. Measuring all bits gives a chance of evading detection of $.75^d$.

Suppose we wish to ensure a probability of no more than $D_a$ of an intruder falsifying an authentication token, and $D_e$ of evading detection of eavesdropping. The protocol has the perhaps unexpected property that more tamper detection bits are required than key bits, if we want to ensure that $D_a$ and $D_e$ are approximately equal. As described above, $D_a = 2^{-k}$ and $D_e = .75^d$. Let $D = D_a = D_e$. Then

$$k = \frac{-\ln D}{\ln 2} \text{ and } d = \frac{\ln D}{\ln .75}$$

so $\frac{d}{k} = 2.41$

Values for $k$ and $d$ needed to implement a required level of security $D$ are

$$k = \left\lceil \frac{-\ln D}{\ln 2} \right\rceil = \lceil -1.44 \ln D \rceil \text{ and}$$

$$d = \lceil -3.48 \ln D \rceil$$

A reasonable level of security for many applications, with $D$ approximately $10^{-6}$, can then be implemented with $k = 17$ and $d = 41$.

### 2.2 Distinguishing tamper detection bits

If Eve can distinguish the tamper detection bits from other bits, she can avoid detection by leaving them undisturbed. However, the location of the tamper detection bits is protected using the symmetric keys shared by Tr and the two parties. Eve would need to decrypt this information in real time, only a few seconds or milliseconds, for it to be useful because it is of no value after Alice and Bob have completed their measurements. Physical means cannot be used to distinguish between entangled and non-entangled bits if Eve has access to only one path (Tr-Alice or Tr-Bob), and thus only one of each pair, because the ability to do so would imply faster than light communication.

### 2.3 Multiple photon splitting

A persistent problem in quantum communication implementations is the difficulty of achieving single photon states. Signals normally contain zero, one, or multiple photons in the same polarization. The *multiple photon splitting*, or *photon number splitting*, attack on quantum protocols involves the eavesdropper deterministically splitting off one photon from each multi-photon signal [Brassard et al., 1999; Huttner et al., 1995]. If Eve measures every single photon and passes along *n*-1 photons undisturbed from each multi-photon state, then her chances of evading detection are increased because the number of tamper detection bits that are effective is reduced to $d' \approx p_1 d$, where $p_1$ is the probability of a single photon state, and the chance of evading detection becomes $0.75^{d'}$. Defending against this attack requires increasing the number of tamper detection bits by a factor of $p_1^{-1}$ to reduce the chance of evading detection to an acceptable level.

### 2.4 Denial of service

The ability to write to or disconnect any channel would allow an attacker to disrupt communication, but this weakness is inherent in any non-redundant communication system. The protocol is therefore suited to networks where channels are assumed to be observable, but cannot be jammed or disconnected.

## 3 Related Work

Zeng and Guo [2000] also describe an authentication protocol based on using entangled pairs. Their protocol uses previously shared secret keys (between each pair of parties) to establish a sequence of measurement bases, and relies on measurement of error rates, as in BB84, to detect the presence of eavesdropping. Jensen and Schack [2000] present a revised version of Barnum's [1999] quantum identification using catalysis. Dusek et al. [1998] combine a classical authentication protocol with quantum key distribution.

## 4 Conclusions and Future Work

This paper describes a protocol for authenticating resources in a network using properties of quantum entanglement. The protocol has a number of advantages over both classical authentication protocols and other quantum protocols. Incorporating conventional symmetric cryptography allows eavesdropping detection to be separated from key distribution, rather than relying strictly on error rates of transmitted keys to detect intrusions. However, an intruder would have only a few seconds or milliseconds to decrypt classically encrypted transmissions between trusted server and workstations.

As described, the protocol relies on idealized properties, and practical implementations may face constraints on transmission efficiency resulting from current technology constraints. The next step required for realization of the protocol is a thorough analysis of effects of these constraints. In particular, multiple rounds of photon distribution between the trusted server and network resources are likely to be required as a result of limits on detection efficiency. Measurements of the efficiency of current implementation schemes, particularly parametric down conversion and weak coherent pulse methods, will be needed.

## 5 Acknowledgements

I am grateful to David Song, Paul E. Black and Ramaswamy Chandramouli for helpful comments and much valuable discussion. I am indebted to Paul for pointing out the limit on bit sampling by Eve. Thanks go to Carl Williams and the NIST QIBEC discussion group for valuable critiques and suggestions.

## 6 References

1. H. Barnum. "Quantum Secure Identification Using Entanglement and Catalysis." LANL archive *quant-ph/*9910072.
2. C.H. Bennett and G. Brassard. Advances in Cryptology: Proceedings of Crypto '84, Springer-Verlag, pp. 475 – 480.
3. Bennett, C. H., Brassard, G., Crepeau, C. and Maurer, U. M., "Generalized Privacy Amplification", *IEEE Transactions on Information Theory*, 1995.
4. G. Brassard, N. Lutkenhaus, T. Mor, and B.C. Sanders, Security Aspects of Practical Quantum Cryptography, *quant-ph*/9911054, 12 Nov. 1999.
5. A. Cabello, Phys. Rev. A 61, 052312 (2000).
6. M. Dusek, O. Haderka, M. Hendrych, R. Myska. "Quantum Identification System", *quant-ph*/9809024, 10 Sept. 1998.
7. S. Felix, N. Gisin, A. Stefanov, H. Zbinden. "Faint Laser Quantum Key Distribution: Eavesdropping Exploiting Multiphoton Pulses," quant-ph/0102062, 12 February, 2001.
8. B. Huttner, N. Imoto, N. Gisin, and T. Mor, Phys. Rev. A. 51, 1863 (1995).
9. J.G. Jensen and R. Schack. "Quantum Authentication and Key Distrubution Using Catalysis", *quant-ph*/0003104, 13 June 2000.
10. D. Song, "Secure Key Distribution by Swapping Quantum Entanglement", quant-ph/0305168.
11. G. Zeng, X. Wang, "Quantum Key Distribution with Authentication", quant-ph/9812022 28 Oct. 1999.
12. G. Zeng, G. Guo, "Quantum Authentication Protocol", *quant-ph*/0001046 13 Jan. 2000.
13. Y-S. Zhang, C-F.. Li, G-C. Guo. "Quantum Authentication Using Entangled State", *quant-ph*/0008044, 16 Aug. 2000.
12. M. Zukowski A. Zeilinger, M.A. Horne, and A.K. Ekert, Phys. Rev. A 71, 4287 (1993).

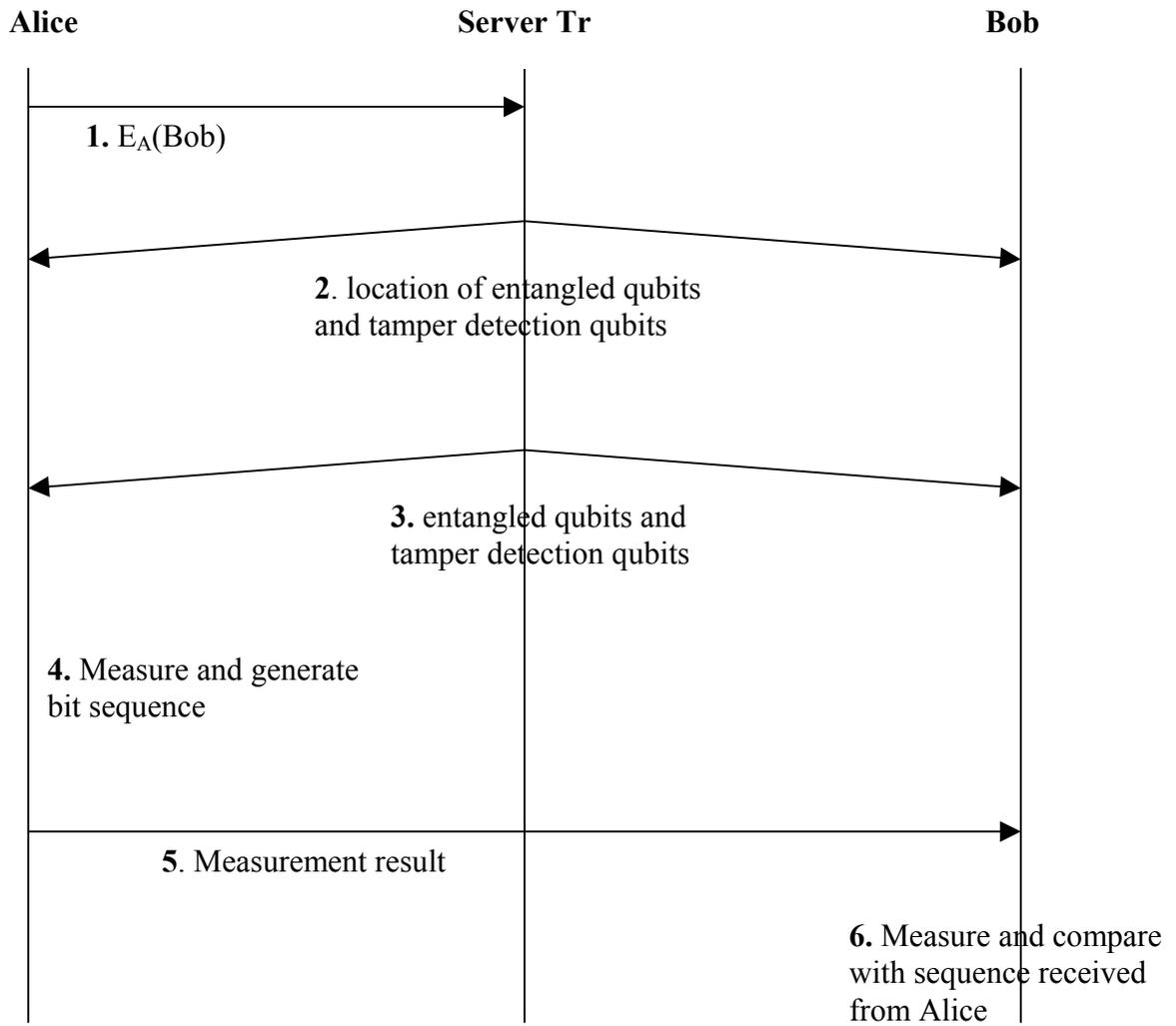

**Figure 1.** Protocol diagram